# Sliding Z Transform: Applications to convolutive blind source separation

XU Peng-fei * , JIA Yin-jie, WANG Zhi-jian

College of Computer and Information, Hohai University, Nanjing 211100, China

**Abstract**: The Z Transform is a mathematical operation in signal processing, which gives a tractable way to solve linear, constant-coefficient difference equations. Based on the classical Z transform and inspired by the thought of sliding DFT, a new definition of Sliding Z Transform(SZT) is introduced and deduced. Then this method is applied to blind source separation, four simulation results are presented to demonstrate its performance when the sliding window $WIN$ is set. It can directly recover time-domain sources from the convolutive mixtures with the help of robust linear mixed blind separation algorithms(such as JADE). It has simple principle and good transplantation capability and can be widely applied in various fields of digital signal processing.

*INTRODUCTION.*—In the theory of discrete signals and systems, Z transform becomes an important mathematical tool. It transforms the mathematical model of the discrete system (the difference equation) to the simple algebraic equation, and simplifies the solution process. Therefore, its position is similar to the Laplace transform in continuous systems and the Z Transform can be considered as a discrete-time equivalent of the Laplace transform. A FIR digital filter has the difference equation:

$$x(n) = s(n) + \sum_{i=1}^{\infty} a_i s(n-i) \quad (1)$$

Here, $s(n)$ is the input signal of the filter, $x(n)$ is the output signal of the filter, $a_i$ is constant value coefficient. The Z transform is applied to the equation, then we have the following.

$$X(z) = A(z)S(z) \quad (2)$$

Here, $S(z)$ is the input signal of the filter, $X(z)$ is the output signal of the filter, $A(z)$ can be considered as a system function.

For the FIR system, the impulse-response is a finite digital sequence, the input and output signals are also finite digital sequences which length is N. so Z transform is applied to the discrete time real value sequence $\{x(k)\}$, then we have the following.

$$X(z) = \sum_{n=0}^{N-1} x(n) z^{-n} \quad (3)$$

*SLIDING Z TRANSFORM.*—Inspired by the thought of sliding DFT[1,2], a concept of sliding Z transform is put forward here. For a data with a whole length of $N$, we add a sliding window (see FIG. 1) with a length of $WIN$, which is mapped to the visible data in this window, and the data outside the window is set to 0. In initialization, the left end of the window is equal to the leftmost end of the data, and the window slides right to the right side of the window until the right end of the window. Up to the right end of the data. In this way, we generate a new set of data. For the sake of understanding, here is a simple example, assuming that the length $N$ of the data $x(n)$ is 8, the length of the sliding window $WIN$ ($WIN = 2^k, k = 0,1,2,3...$) is 4, the step of sliding window $P = 1$ ($1 \leq P \leq WIN$), and then we take the time $n=0,1,2,3,4$ ($0 \leq n \leq N - WIN$).

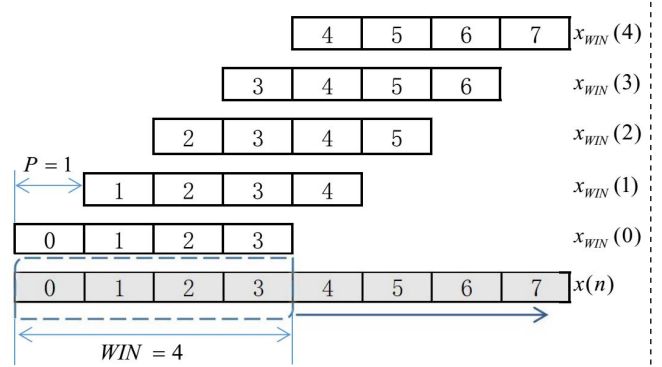

FIG. 1. Sliding window (P=1)

The data in these time windows are:

$$x_{WIN}(0) = \{x(0), x(1), x(2), x(3)\}$$
$$x_{WIN}(1) = \{x(1), x(2), x(3), x(4)\}$$
$$x_{WIN}(2) = \{x(2), x(3), x(4), x(5)\}$$
$$x_{WIN}(3) = \{x(3), x(4), x(5), x(6)\}$$
$$x_{WIN}(4) = \{x(4), x(5), x(6), x(7)\}$$

These new generated data are recorded as $x_{WIN}(0)$、$x_{WIN}(1)$、$x_{WIN}(2)$、$x_{WIN}(3)$ and $x_{WIN}(4)$. These data are sent to the Z converter in turn. We get $X(z,0)$ (see FIG. 2)、$X(z,1)$、$X(z,2)$、$X(z,3)$ and $X(z,4)$.

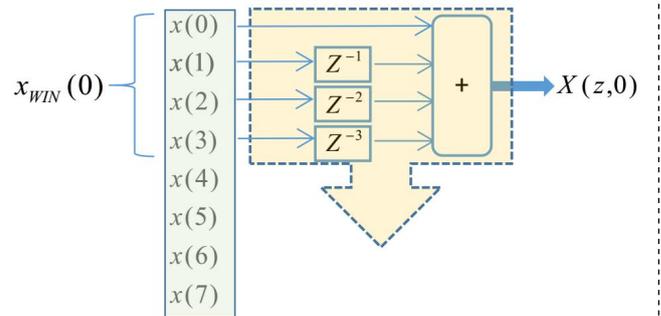

FIG. 2. Sliding Z transform at time 0

From the above description, $S(z)$ and $X(z)$ can be written as the form of sliding window Z transform.

$$S(z,n) = \sum_{m=0}^{WIN-1} s(n+m)z^{-m} \quad (4)$$

$$X(z,n) = \sum_{m=0}^{WIN-1} x(n+m)z^{-m} \quad (5)$$

Therefore, the formulas (6) and (7) are derived from the formula (4) and (5) at n+1 time as follows.

$$S(z,n+1) = \sum_{m=0}^{WIN-1} s(n+m+1)z^{-m} \quad (6)$$

$$X(z,n+1) = \sum_{m=0}^{WIN-1} x(n+m+1)z^{-m} \quad (7)$$

The formula (6) and (7) are multiplied by $Z^{-1}$:

$$S(z,n+1)*Z^{-1} = \sum_{m=0}^{WIN-1} s(n+m+1)z^{-(m+1)} \quad (8)$$

$$X(z,n+1)*Z^{-1} = \sum_{m=0}^{WIN-1} x(n+m+1)z^{-(m+1)} \quad (9)$$

If you subtract formula(8) from formula(4), you get formula(10).

If you subtract formula(9) from formula(5), you get formula(11).

$$S(z,n) - S(z,n+1)*Z^{-1} = s(n) - s(n+WIN)Z^{-WIN} \quad (10)$$

$$X(z,n) - X(z,n+1)*Z^{-1} = x(n) - x(n+WIN)Z^{-WIN} \quad (11)$$

*APPLICATION AND SIMULATION.*—We regard the data $S(z,n) - S(z,n+1)*Z^{-1}$ in formula (10) as the new input signal $S'(n)$ in Z domain and the data $X(z,n) - X(z,n+1)*Z^{-1}$ in formula (11) as the new output signal $X'(n)$ in Z domain. In the problem of blind signal separation[3], the formula (1) can be considered as a convolutive mixture (see FIG. 3).

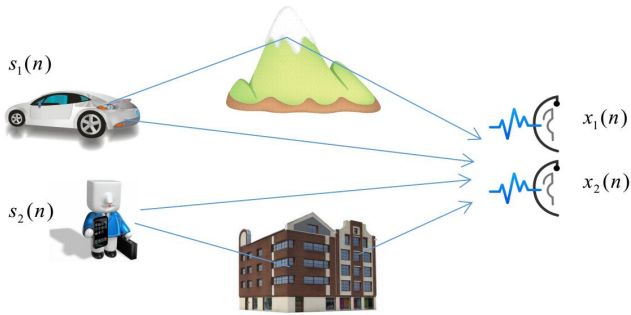

FIG. 3. Convolutive mixture

The blind separation of convolutive mixture signals is a nodus in blind source separation. The formula (2) can be regarded as instantaneous mixture. It is known by the nature of the Z transformation. Convolutive mixed signals in the time domain have become instantaneous linear mixed signals in the Z domain. In this way, we can separate the source signal $S'(n)$ from the Z domain by using the mature linear mixed blind separation algorithm for $X'(n)$.

Then rests on $S'(n) = s(n) - s(n+WIN)Z^{-WIN}$, namely, $s(n+WIN) = (s(n) - S'(n))*Z^{WIN}$, therefore, the source signal $s(n)$ in the time domain is further obtained. For the received $i$ ($i \geq 1$) mixed signals $x_i(n)$, there is a flow chart of blind signal separation of convolutive mixtures based on Sliding Z Transform(SZT) as follows.

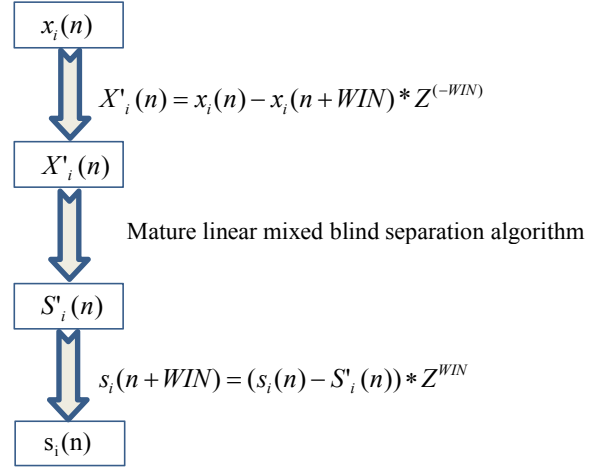

FIG. 4. Flow chart of blind signal separation of convolutive mixtures based on SZT

It should be noted that $Z$ is a random number generated by the system, it needs to satisfy the condition $0 < Z < 1$ to make the system stable.

Here four Matlab simulations are presented to demonstrate the performance of the sliding Z transform. The mixing matrix $A$ is generated randomly from the standard uniform distribution on the open interval (0,1). The number of paths is set to 2, then the convolutive mixed signal is obtained by using the formula (1) and Fig.3. The two mixing matrixs are $A' = \begin{bmatrix} a'_{11} & a'_{12} \\ a'_{21} & a'_{22} \end{bmatrix}$ and

$A'' = \begin{bmatrix} a''_{11} & a''_{12} \\ a''_{21} & a''_{22} \end{bmatrix}$,

$x_1(n) = a'_{11}s_1(n) + a''_{11}s_1(n-1) + a'_{12}s_2(n) + a''_{12}s_2(n-1)$,
$x_2(n) = a'_{21}s_1(n) + a''_{21}s_1(n-1) + a'_{22}s_2(n) + a''_{22}s_2(n-1)$.
The steps of the algorithm described in Fig. 4 are used to separate these mixed signals $x_1(n)$ and $x_2(n)$. Many linear mixed blind separation algorithms can be used to recover the sources. Among them, the robust JADE(Joint Approximate Diagonalization of Eigenmatrices) algorithm[4] is selected to separate the new output signal $X'(n)$. Evaluating the performance of blind source separation, a correlation coefficient $C$ is introduced as a performance index[5].

$$C(x,y) = \frac{\text{cov}(x,y)}{\sqrt{\text{cov}(x,x)}\sqrt{\text{cov}(y,y)}} \quad (12)$$

$C(x,y) = 0$ means that x and y are uncorrelated, and the signals correlation increases as $C(x,y)$ approaches unity,

the signals become fully correlated as $C(x,y)$ becomes unity.

In the first experiment, the sources are two clear sound signals of a bird and a creek with 2000000 samples and 44100Hz sample rates(downloaded from http://www.orangefreesounds.com/category/sound-effects/). Letting $WIN = 8$.

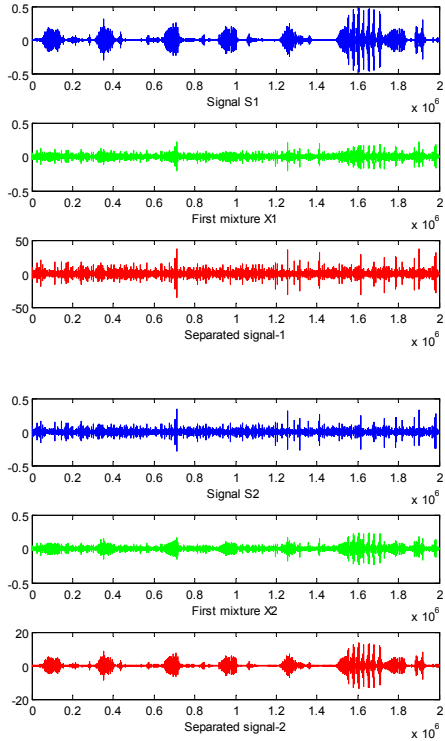

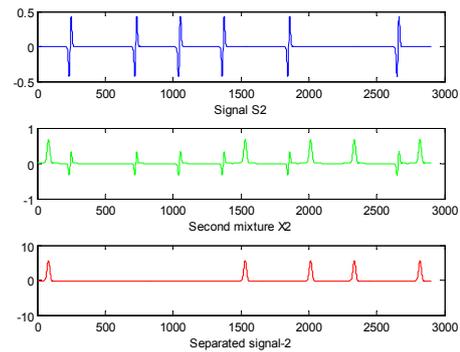

FIG. 6. The time-domain waveforms of sparse signals after blind source separation

After separation,the correlation coefficients between the separated signals and the sources are 0.9987 and 0.9965 respectively,as shown in Fig.6. The sources are well recovered.The replacement of other sparse signals is still valid.

In the third experiment, the sources are two mobile communication signals (Represented by QPSK signals with different carriers) with 200000 samples. These signals pass through the AWGN (Additive White Gaussian Noise) channel which is the most basic model of noise and interference. SNR(Signal to Noise Ratio) is set to 20 dB. Letting $WIN = 64$.

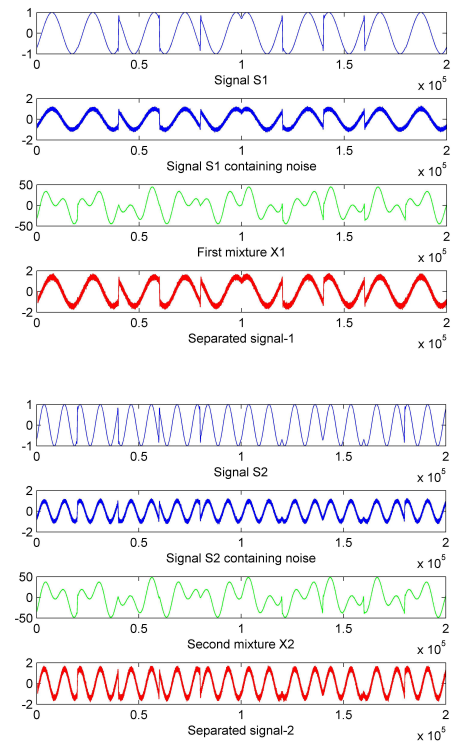

FIG. 5. The time-domain waveforms of sound signals after blind source separation

After separation, the correlation coefficients between the separated signals and the sources are 0.9988 and 0.9943 respectively,as shown in Fig.5. The sources are well recovered. The replacement of other sound signals is still valid.

In the second experiment, the sources are two sparse signals[6] (Represented by Gauss pulse signals of first-order and second-order) with 3000 samples. Letting $WIN = 8$.

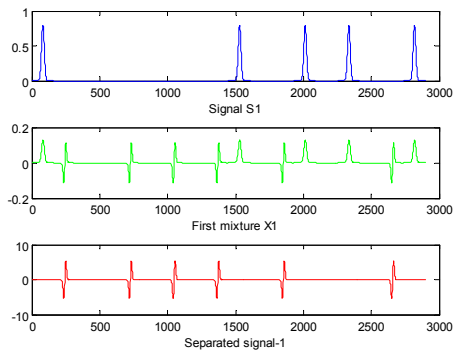

FIG. 7. The time-domain waveforms of mobile communication signals after blind source separation

After separation,the correlation coefficients between the separated signals and the sources are 0.9934 and 0.9860 respectively,as shown in Fig.7. the sources are well recovered. After separating the data, we can de-noise the separated signals[7] and get the more pure signal before proceeding the demodulation. The replacement of other mobile communication signals with different or same carriers[8] is still valid.

In the fourth experiment, we increase the noise energy to make $SNR < 0$, the source signal completely "submerged" in the noise[9]. The sources are signals of QPSK and WGN (White Gaussian Noise) [10] with 200000 samples. Letting $WIN = 512$.

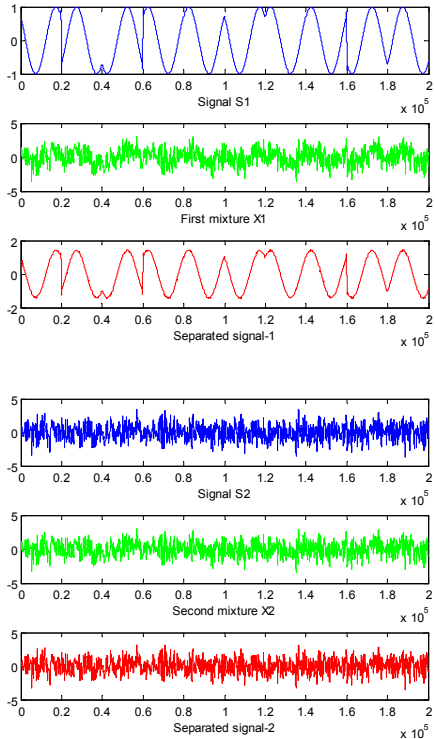

FIG. 8. The time-domain waveforms of signals in very low SNR environments after blind source separation

After separation, the correlation coefficients between the separated signals and the sources are 0.9987 and 0.9944 respectively, as shown in Fig.8. The sources are well recovered. The replacement of other mobile communication signals with different or same carriers is still valid.

These four experiments above demonstrate that convolutive signals can be separated using sliding Z transform. More importantly, the permutation indeterminacy and amplitude indeterminacy will not affect the final separation performance[11]. It is worth mentioning that when the number of paths is set to 1, the convolutive mixing becomes instantaneous mixing, the mixed signal $x_1(n) = a'_{11} s_1(n) + a'_{12} s_2(n)$ and $x_2(n) = a'_{21} s_1(n) + a'_{22} s_2(n)$. These experiments are done again, after separation, the correlation coefficients between the separated signals and the sources are 1.0000 and 1.0000 respectively, the separation effect is perfect.

*CONCLUSION.*—In this paper, we have presented a single step ($P = 1$) Sliding Z Transform(SZT) and applied it to the blind separation problem. we looks into the possibility of its application, experiments on four different scenes demonstrate its fine performance. It can directly recover time-domain sources from the convolutive mixtures with the help of JADE algorithms. Due to its simple principle and good transplantation capability, it can be widely applied in various fields of digital signal processing such as image enhancement and restoration of images, biomedical signal processing(such as ECG,EEG,EMG,EOG,EGG) and array signal processing. Some questions that need further study include the influence of the sliding window size on the blind separation system, multi-step($P \geq 2$) sliding Z transform and adaptive sliding Z transform and so on.

*xpf@hhu.edu.cn

[1] Jacobsen E, Lyons R. The sliding DFT[J]. Signal Processing Magazine IEEE, 2003, 20(2):74-80.
[2] Jacobsen E, Lyons R. An update to the sliding DFT[J]. IEEE Signal Processing Magazine, 2004, 21(1):110-111.
[3] J. H´erault, C. Jutten, and B. Ans, "D´etection de grandeurs primitives dans un message composite par une architecture de calcul neuromim´etique en apprentissage non supervis´e," in Proc. GRETSI, Nice, France, 1985, pp. 1017–1020.
[4] Cardoso J F, Souloumiac A. Blind beamforming for non-Gaussian signals[C]//IEE proceedings F (radar and signal processing). IET Digital Library, 1993, 140(6): 362-370.
[5] Peled R, Braun S, Zacksenhouse M. A blind deconvolution separation of multiple sources, with application to bearing diagnostics[J]. Mechanical Systems & Signal Processing, 2005, 19(6):1181-1195.
[6] Rossi P V, Kabashima Y, Inoue J. Bayesian online compressed sensing[J]. Physical Review E, 2016, 94(2-1):022137.
[7] Torresforné A, Marquina A, Font J A, et al. Denoising of gravitational wave signals via dictionary learning algorithms[J]. Physical Review D, 2016, 94(12).
[8] XU Peng-fei, LIU Nai-an, FU Wei-hong. Convolutive blind source separation applied to the communication signals that with same carrier frequencies and modulation[J]. Journal of Chongqing University of Posts and Telecommunications (Natural Science Edition), 2010, 22(3):312-316.(in Chinese)
[9] Knollmüller J, Enßlin T A. Noisy independent component analysis of autocorrelated components[J]. Phys.rev.e, 2017, 96(4).
[10] Dechant A, Baule A, Sasa S I. Gaussian white noise as a resource for work extraction[J]. Phys.rev.e, 2017, 95(3-1):032132.
[11] ZHANG Gui-bao, LI Jia-wen, LI Cong-xin. A novel blind deconvolution algorithm using single frequency bin [J]. Journal of Zhejiang University SCIENCE A, 2007,8(8): 1271-1276.